\documentclass[preprint]{aastex}

\usepackage[center]{caption}

\newcommand{\el}{{\rm e}}

\newcommand{\h}{{\rm H}}

\newcommand{\nh}{n_{\rm H}}
\newcommand{\Nh}{N_{\rm H}}
\newcommand{\hp}{{\rm H}^+}

\newcommand{\hep}{{\rm He}^+}

\newcommand{\hm}{{\rm H}_2}
\newcommand{\xhm}{x({\rm H}_2)}

\newcommand{\hmin}{{\rm H}^-}

\newcommand{\oh}{{\rm OH}}

\newcommand{\ohp}{{\rm OH}^+}

\newcommand{\htwoo}{{\rm H}_{2}{\rm O}}
\newcommand{\xhtwoo}{x({\rm H}_{2}{\rm O})}

\newcommand{\htwoop}{{\rm H}_{2}{\rm O}^{+}}

\newcommand{\co}{{\rm CO}}

\newcommand{\alphah}{\alpha_{\rm h}}
\newcommand{\Td}{T_{\rm d}}
\newcommand{\Tg}{T_{\rm g}}
\newcommand{\nd}{n_{\rm d}}
\newcommand{\rhod}{\rho_{\rm d}}
\newcommand{\rhog}{\rho_{\rm g}}

\newcommand{\Tx}{T_{\rm X}}
\newcommand{\Lx}{L_{\rm X}}
\newcommand{\percc}{\rm \,cm^{-3}}

\newcommand{\persqcm}{\rm \,cm^{-2}}

\newcommand{\gpercc}{\rm \,g\,cm^{-3}}
\newcommand{\ps}{{\rm s}^{-1}}
\newcommand{\ccps}{{\rm cm}^{3} {\rm s}^{-1}}

\newcommand{\ergps}{{\rm erg}\,{\rm s}^{-1}}

\def\micron{\hbox{$\mu$m}}
\newcommand{\bcen}{\begin{center}}
\newcommand{\ecen}{\end{center}}
\newcommand{\be}{\begin{equation}}
\newcommand{\ee}{\end{equation}}
\newcommand{\bdis}{\begin{displaymath}}
\newcommand{\edis}{\end{displaymath}}

\def\noi{{\noindent}}
\def\ra{{\rightarrow}}

\begin{document}

\title{Formation of Water in the Warm Atmospheres of Protoplanetary Disks}

\author{A.E. Glassgold\altaffilmark{1}, R.
  Meijerink\altaffilmark{1,2}, and J.R. Najita\altaffilmark{3}}

\altaffiltext{1}{Astronomy
  Department, University of California, Berkeley, CA 94720, United
  States}\email{aglassgold@astro.berkeley.edu}

\altaffiltext{2}{California Institute of Technology, Division of
  Geological and Planetary Sciences, MS 150-21, Pasadena, CA 91125}
\email{rowin@gps.caltech.edu}

\altaffiltext{3}{National Optical Astronomy Observatory, 950 North
  Cherry Avenue, Tucson, AZ 85719}
\email{najita@noao.edu}

\begin{abstract}

The gas-phase chemistry of water in protoplanetary disks is analyzed 
with a model based on X-ray heating and ionization of the disk atmosphere.  
Several uncertain processes appear to play critical roles in generating 
the column densities of warm water that are detected from disks at 
infrared wavelengths. The dominant factors are the 
reactions that form molecular hydrogen, including 
formation on warm grains, and the ionization and heating of the 
atmosphere. All of these can work together to produce a region of high 
water abundances in the molecular transition layer of the inner disk 
atmosphere, where atoms are transformed into molecules, the temperature 
drops from thousands to hundreds of Kelvins, and the ionization begins 
to be dominated by the heavy elements. Grain formation of molecular 
hydrogen and mechanical heating of the atmosphere can play important roles 
in this region and directly affect the amount of warm water in 
protoplanetary disk atmospheres. Thus it may be possible to account 
for the existing measurements of water emission from Tauri disks without 
invoking transport of water from cooler to warmer regions. 
The hydroxyl radical OH is under-abundant in this model of disk 
atmospheres and requires consideration of additional production and 
excitation processes.  
\end{abstract}


\section{Introduction}

The water molecule plays a significant role in the evolution of
protoplanetary disks and in the formation of small and large bodies
including planets (e.g., Ciesla \& Cuzzi 2006; Jewitt et al.~2006).
Recent results from the Spitzer Space Telescope ({\it Spitzer}) reveal
rich mid infrared (MIR) water emission spectra from 10-35 microns that
appear to be a common characteristic of T Tauri disks (Carr \& Najita
2008; Salyk et al.~2008).  Assuming thermal equilibrium populations,
typical temperatures and column densities have been estimated to be $T
\sim 500-1000$\,K and $N(\htwoo) \sim 10^{17}-10^{18} \persqcm$, with
emitting areas of the order of a few AU in radius.   Mid infrared 
emission from other molecules (HCN,
C$_2$H$_2$, OH, CO$_2$) is also detected with similar temperatures and
likely similar emitting radii, e.g., Carr and Najita (2008).  These
MIR lines appear to probe lower gas temperatures and larger disk radii
than the previously known suite of gaseous inner disk diagnostics.  UV
fluorescent emission from $\hm$ and near-infrared transitions of CO
probe warm gas temperatures (1000-3000\,K) and primarily the inner
$ < 1$ AU of the disk (Najita et al.~2007). Measurements of 
molecular line emission have been obtained for many other protoplanetary 
disks by {\it Spitzer} and should become available for analysis and
interpretation in the near future. 

Our earlier model of the thermal-chemical structure of an X-ray
irradiated generic T Tauri disk atmosphere (Glassgold et al.~2004;
henceforth GNI04) included the synthesis of $\htwoo$ by 
temperature-sensitive neutral radical reactions (Table 1 of that paper),
but it did not predict strong water emission.  Possible explanations 
for this deficiency might be that water is dredged up from deeper parts 
of the disk atmosphere by turbulent mixing, or that it is transported 
radially by inward migration of intermediate-sized bodies and desorbed
(Ciesla \& Cuzzi 2006; Najita et al.~2007; Salyk et al.~2008). In this case, 
measurements of water abundances in disk atmospheres might be used to 
trace dynamical processes such as turbulence and migration. However, 
disk gas phase chemistry is affected by several often poorly known 
physical processes, so it is important to fully explore non-dynamical 
explanations. In this paper, we analyze some of these processes in 
light of the recent {\it Spitzer} $\htwoo$ observations 
by exploring alternative or modified chemical pathways that may 
enhance the abundance of water and other molecules in disk atmospheres.
A longer term goal is to attempt to use observations of water to infer 
the physical properties of protoplanetary disks.  

This study complements recent work on disk chemistry by focusing on the 
thermal-chemical properties of an X-ray irradiated disk atmosphere that 
has experienced grain growth and settling. 
X-ray induced chemistry has a long history going back 25 years 
(Krolik and Kallman 1983). References to early work is given in the 
pioneering papers by Lepp and Dalgarno (1996) and Maloney et al.~(1996). 
The latter calculate the water abundance in X-ray-transition 
regions, as do Meijerink and Spaans (2005). Stauber et al.~(2005)
develop X-ray chemistry for spherically symmetric envelopes around 
YSOs with specific emphasis on water (Stauber et al.~2006).

Many disk chemical models 
consider the outer region of disks, whereas we concentrate on the inner 
$< 10-20$\,AU.  Several recent papers treat inner disk gas phase chemistry
(Markwick et al.~2002; Nomura and Millar 2005; Nomura et al.~2007, 2009;
Ag\'{u}ndez et al.~2008; Gorti and Hollenbach 2008; Woods and Willacy 2009).
Nomura et al.~(2007) focus on $\hm$, and Nomura et al.~(2009) follow 
molecules along prescribed stream lines and find that the abundances 
of many organic molecules change as they enter the inner disk region, 
notably methanol but not water. Markwick et al.~and Ag\'{u}ndez et al.~include 
X-ray ionization in terms of a phenomenological ionization rate, whereas 
Woods and Willacy use a more complete theory based on Gorti and 
Hollenbach (2004). Ag\'{u}ndez et al.~carry out time-dependent abundance 
calculations, in some cases at fixed temperatures. We shall discuss the 
water abundances obtained by these authors in the Discussion section below.

In this paper we analyze a number of physical processes that can
affect the abundance of water in the warm part of the disk atmosphere
($ T > 200-300$\,K) using the GNI04 model. In Sec.~2, we discuss the 
role played by dust through the thermal coupling of the gas
and dust; the efficiency of the gas-phase reactions that produce $\hm$
and $\htwoo$; the role of $\hm$ and $\htwoo$ formation on dust grains;
and finally the effects of enhanced mechanical heating of the
atmosphere.  Implementation of these processes with our thermal-chemical
model, along with new information about their efficiency, indicates
that column densities of warm water in the range observed can be
obtained without invoking radial or vertical transport. The
implications of this results are taken up in Sec.~3 and in the
Conclusions in Sec.~4.

% Section 2
\section{Thermal-Chemical Model of X-ray Irradiated Disks}

% Sec. 2.1
\subsection{The Basic Model}
This work builds on our earlier model of the thermal-chemical
structure of the inner disk of a T Tauri star (GNI04; see Meijerink
et al.~2008 for corrections and additional applications).  The
disk is irradiated by stellar X-rays which ionize and heat the disk
surface. 
The X-ray luminosity is $\Lx = 2 \times 10^{30} \ergps$;
a thermal spectrum is used with $\Tx = 1.0$\,keV and a low-energy 
cutoff of $E_0 = 200$\,eV. 
As in GNI04 and our earlier work, cosmic rays are ignored on 
the basis of their being blown away from the inner disk by the
stellar wind. 
The total gas density distribution and the dust temperature 
are taken from D'Alessio et al.~(1999, 2001), and the gas-phase 
temperature and abundances are solved for self-consistently keeping 
the density and dust temperature fixed. In a more general model 
that treats the dust and gas separately, a new equilibrium would 
be calculated, including the effects of changing the gas temperature 
on the hydrostatic balance and on the dust temperature.

The gas temperature in the current model is determined by balancing 
heating from X-rays and mechanical processes with cooling by line 
radiation.  GNI04 employed a simple chemistry with 25 species and 125 
reactions that included the heavy atoms, C and O, as well as H and He. 
In order to provide a more complete theory of X-ray ionization, the 
chemical code used here treats all of the most abundant heavy atoms 
and includes nearly 400 reactions (Meijerink \& Glassgold 2009). As 
mentioned in the Introduction, chemical synthesis is achieved by 
neutral radical 
reactions with moderate barriers that can be overcome in the warm upper 
atmosphere of the disk. Ionic reactions also contribute, but their 
most important role is to destroy molecules. One difference with GNI04 is  
that we use depleted abundances, defined relative to hydrogen (nuclei) 
as usual, with $x_{\rm O} = 3.5\times 10^{-4}$ and 
$x_{\rm C} = 1.4 \times10^{-4}$. The thermal part of the program 
is unchanged from GNI04 even as changes are made in the chemistry.
For example, NeII and water cooling are still ignored, although 
their inclusion could change the quantitative aspects of the present 
results.  In addition, a significant fraction of the stellar NIR 
flux may be captured by water molecules.  A consistent theory of 
the thermal effects of water in disk atmospheres requires 
consideration of both heating and cooling. We postpone treating this 
important topic to a future study. Also omitted in this pure 
X-ray irradiated disk model are the effects of UV radiation, which 
can affect the abundances of OH, $\htwoo$ and other species. 
Efforts are under way to include both UV radiation and X-rays in 
our models.

Dust grains play a {\it direct} role in the thermal balance via the
familiar collisional coupling of the dust and the gas. Cooling of the
gas by the dust becomes important in the region we refer to as the 
{\it transition region} where the
temperature and ionization level are rapidly decreasing and atomic
species are transformed into molecules. Since the chemistry is sensitive
to temperature, the dust {\it indirectly} affects the chemical
abundances.  In the GNI04 model, it was assumed that the dust in the
inner disk had experienced growth and settling. The grain area per H
nucleus was given by (Eq. A4 of that paper),
\be
\label{grain_area}
\frac{\nd}{\nh} <a^2> =  
\frac{\rhod / \rhog}{a_{\rm geom}}
\frac{m}{(4\pi/3)\tilde{\rho}}, 
\hspace{0.10in} a_{\rm geom} = (a_1a_2)^{1/2},
\ee
where $\nd$ and $\rhod$ are the volumetric number and mass density of
the dust grains, $\rhog$ is the mass density of gas, 
$\tilde{\rho}$ is the internal density of a typical
grain, $m$ is the mean mass of the gas per H nucleus,
and $a_1$ and $a_2$ are the minimum and maximum grain radii for a MRN
distribution.  The key parameter in Eq.~\ref{grain_area} is the dust to
gas ratio $\rhod / \rhog$ divided by $a_{\rm geom}$, the geometric
mean of $a_1$ and $a_2$. In GNI04, these parameters were set at $\rhod /
\rhog=0.01$ and $a_{\rm geom} = 0.707$\,$\mu$m, following Wood et al.~(2001).  This choice corresponds to a factor of 20 increase in
$a_{\rm geom}$ and a factor of 20 decrease in effective grain surface
area relative to the interstellar medium. Wood et al.~actually used a
MRN distribution with $a_1= 0.01$\,$\mu$m, an exponential cutoff at $a
= 50\,\micron $ and $a_2 = 1$\,mm. 
Had we used the grain model in
D'Alessio et al.~(2001; $a_1= 0.005$\,$\mu$m and $a_2 = 1$\,mm), then
$a_{\rm geom}$ would be $2.24$\,$\mu$m, or 3.2 times larger than in GNI04. 
In this paper we follow GNI04 but also consider larger values of $a_{\rm geom}$, i.e., $a_{\rm geom} = 7.07$\,$\mu$m as well as
$a_{\rm geom}= 0.707$\,$\mu$m.  This range gives grain areas per H
nucleus that are roughly consistent with the dust depletion deduced by Furlan
et al.~(2006) for the upper atmospheres of T Tauri disks observed with
{\it Spitzer}.

% Sec. 2.2
\subsection{The Dependence of the Water Abundance on the Physical
Properties}

Only gas phase reactions were considered by GNI04 for the
formation of H$_2$ because the grains in disk
atmospheres were considered too warm for the usual grain synthesis
familiar from the interstellar medium (e.g., Hollenbach \& Salpeter 1971) 
to be efficient. Instead, H$_2$ was formed by radiative attachment 
of electrons
\be
\label{hminusreacs}
\el + \h \ra \hmin + h\nu \hspace{0.5in} 
\hmin + \h \ra \hm + \el,
\ee
and by three-body reactions, 
\be
\label{3bodyreacs}
3\h \ra \hm + \h 	\hspace{0.5in} 2\h + \hm \ra 2\hm,
\ee
the latter accompanied by collisional dissociation,
\be
\h + \hm \ra 3\h	\hspace{0.5in} \hm + \hm \ra \hm + 2\h. 		
\ee     
These gas phase reactions led to H$_2$ abundances approaching $\sim 10^{-3}$  
in the surface region, a level sufficient to produce significant amounts 
of warm CO ($ \sim 10^{17} - 10^{18} \persqcm$) by a sequence of neutral 
radical reactions that begins with the slightly endothermic reaction 
\be
\label{slowestneutral}
{\rm O} + \hm \ra \h + {\rm OH} 
\ee 
and leads to water by the exothermic reaction,
\be
\label{ohneutral}
{\rm OH} + {\rm H}_2 \ra \htwoo + {\rm H}.		
\ee 
Both of these reactions have barriers that require warm conditions 
to be effective, i.e., $T \gtrsim 200-300$\,K. OH also leads to O$_2$
and CO by fast exothermic radical reactions without significant barriers,
\be
{\rm OH} + {\rm O} \ra {\rm O}_2 + {\rm H} \hspace{0.5in}
{\rm C } + {\rm OH} \ra {\rm CO} + {\rm H}	.
\ee
OH and $\htwoo$ are mainly destroyed by charge transfer with $\hp$,
\be
\label{ohionic}
\hp + {\rm OH}  \ra \ohp + {\rm H},		
\ee 
\be
\label{h2oionic}
\hp + \htwoo  \ra \htwoop + {\rm H}.		
\ee 

The chemistry defined by the above equations starting with Eq.~5 
can be approximated by the following formula for the abundance of water,
\be
\label{closed_formula}
\frac{x(\htwoo)}{x(\rm O)} \sim \kappa(T) 
\left (\frac{x(\hm)}{x(\hp)}\right)^2
\ee
where
\be 
\kappa \equiv \frac{k_5k_6}{k_8k_9} \thickapprox 10^{-18}T^{4.27} \exp(-4823/T).
\ee
The ratio of water to atomic oxygen depends {\it quadratically} 
on the ratio of $\hm$ to $\hp$ because the synthesis of water 
proceeds via {\it two} radical reactions with $\hm$, (Eqs.~5 and 6), 
and because the precursor radical OH as well as H$_2$O are destroyed 
primarily by fast charge exchange with $\hp$ (Eqs.~8 and 9). The 
factor $\kappa(T)$ is sensitive to temperature because the product 
of the rate coefficients, $k_5$ and $k_6$, for the two synthesis 
reactions is temperature-sensitive, whereas the product of the two 
rate coefficients, $k_8$ and $k_9$, for the two ionic destruction 
reactions is not. Equation 10 ignores several factors that can limit its 
accuracy: the role of the backward reactions to Eqs.~5 and 6 at high 
temperatures; the contribution of ion-molecule reactions to the 
synthesis of water; and the destruction of OH and H$_2$O by He$^+$ ions. 

The main purpose of introducing Eq.~10 is to elucidate the underlying 
physics of the molecular transition and not to reproduce the full
model calculations. The main conclusion from this approximate treatment 
is that the water abundance in disk atmospheres is sensitive to two 
quantities that vary rapidly in the transition zone between the very 
warm upper atmosphere and the cool near-midplane region, the temperature 
and the ratio of the molecular hydrogen abundance to the H$^+$ abundance. 
Any process that affects these quantities can change the abundance of 
water. For our model of an X-ray irradiated disk, the X-rays 
play a key role since they heat and ionize the gas.

The GNI04 model calculations yielded small column densities of 
warm water ($\lesssim 10^{13} \persqcm$ at $T > 200-300$\,K).  In contrast, 
observations indicate the presence of abundant warm water, e.g., 
Carr et al.~(2008) report $N(\htwoo) = 6.5 \times 10^{17} \persqcm$ 
at $T=500-600$\,K in AA Tau, and Salyk et al.~find similar columns in 
AS 205A and DR Tau. According to our 
calculations, the water is probably located in the vertical 
thermal-chemical transition region that occurs between the hot and 
cool parts of the disk, where the gas temperature is relatively high, 
e.g., in the 300-2000\,K range.  The observed water 
lines are in emission, which further suggests that the lines 
form above the MIR dust photosphere. The deficiency of the
GNI04 model in this respect suggests three possibilities: (1) other 
processes heat the atmosphere over larger column densities, so that 
the water deeper down in the atmosphere is put into emission; 
(2) water is transported dynamically and mixed into the upper 
atmosphere where it is heated; (3) more efficient chemical 
pathways form water in the disk atmosphere. We mainly focus on 
options (1) and (3) in this paper.  

% Sec. 2.3
\subsection{Gas Phase Reactions}

Reaction~\ref{slowestneutral} is an important destruction mechanism 
for $\hm$. The recent critical review by Baulch et al.~(2005) contains 
laboratory measurements of this reaction from 300-3500\,K. 
The low-temperature data ($T\gtrsim 300$\,K) are 20 years old and were 
obtained with shock tubes or flash photolysis.  GNI04 used the value 
recommended in the earlier Baulch et al.~(1992) review, 
\be
\label{baulcheq}
k_5 = 8.5\times 10^{-20}\,T^{2.67}\, e^{-3163/T} \ccps.
\ee
The more recent experiments all refer to the higher-temperature 
regime, $T =1500-3500$\,K, and the above fit is still satisfactory
at lower temperatures. Nonetheless, considerable uncertainty 
must be attached to Eq.~\ref{slowestneutral} ($\sim 50-100\%$), and 
even more when it is extrapolated below 300\,K. GNI04 mistakenly
used the exponent 1.67 instead of 2.67, seriously underestimating 
the production of oxygen molecules while underestimating the 
destruction of $\hm$ via Eq.~\ref{slowestneutral}. 
This error was corrected in MGN08, but even with this change only 
a small amount of warm 
water was generated, e.g., $N(\htwoo) \lesssim 10^{13}\,\persqcm$ 
at 1 AU.  Significant water abundances, say $x(\htwoo) > 10^{-6}$, 
were only achieved at large vertical column densities 
($\Nh > 10^{22}\,\persqcm$) where the dust temperature is barely 
above freeze-out.  This situation is illustrated by the dotted curves
in the middle and bottom panels of Figure 1 for a radial distance of 1\,AU. 

Another critical gas phase process is three-body formation of  
$\hm$, the reactions in Eq.~\ref{3bodyreacs}.
In the past, these three-body rate coefficients were
obtained with detailed balance from measurements of the inverse 
reactions, usually those of Jacobs et al.~(1967). Flower and Harris 
(2007) have used new equilibrium constants to obtain three-body
formation rate coefficients that are 
significantly larger than those given in the review of 
Cohen and Westberg (1983) and used by GNI04. For example, the first 
forward reaction in Eq.~3 is, as given by Flower and Harris,
\be
\label{threebody}
k_3 (\h \ra \hm + \h )  = 1.44 \times 10^{-26} T^{-1.54}\,{\rm cm}^6 \ps,
\ee
which is to be compared with the Cohen and Westberg rate, 
$8 \times 10^{-33}$\, cm$^{6}$ s$^{-1}$.

The top panel of Figure 1 shows the change in the abundances of 
hydrogenic species obtained using the new rate coefficients. 
The difference between old and new rate coefficients is not
manifest until $\Nh > 5 \times 10^{20} \persqcm$ because the
$\hm$ formation proceeds mainly via the H$^-$ route, 
Eq.~\ref{hminusreacs}, at high altitudes.The strong rise in the $\hm$ abundance as
$\Nh$ approaches $5 \times 10^{20} \persqcm$ occurs because the 
three-body rate increases rapidly with decreasing temperature
in a region where the atomic H density is also increasing towards 
$10^{10} \percc$. As a result, the H-$\hm$ transition to be 
is shifted to smaller vertical column densities, which means to 
warmer regions higher in the atmosphere.  
As can be seen Figure 1, this also leads to a significant increase 
in the amount of warm water,
aided by a drop in the abundance of $\hp$ according to 
Eqs.~\ref{ohionic} and \ref{h2oionic}.
The incomplete transitions of atomic O to OH 
and $\htwoo$ (to abundances $\sim 10^{-6}$) now occur in the 
same general region as the hydrogen and carbon transitions. 
The solid curves in Figure 1 define the first in a series of 
comparison models to be introduced in this section (''model 1"). 
The other models are defined in Table 1; their specific properties 
are given later in Table 2. Note that the parameter $a_{\rm geom}$ 
listed in Table 1 determines the dust cooling of the gas as well 
as the formation of $\hm$ on grains.

\begin{center}
  \begin{table}
    \caption{Chemical Models}
    \begin{center}
      \begin{tabular}{lllcl}    
        \hline
        \hline
        Model & Chemistry & $a_{\rm geom}$ [$\mu$m] & $\alpha_{\rm h}$\\        
        \hline
        1	&	pure gas phase	 		& 0.707		& 0.01	\\ 
        2	&	$\hm$ formation on grains	& 0.707		& 0.01	\\
        3	&	$\hm$ formation on grains	& 7.070		& 0.01	\\
        4	&	$\hm$ \& $\htwoo$ formation on grains	& 0.707	& 0.01	\\
        5	&	pure gas phase	 		& 0.707		& 1.00	\\
        6	&	$\hm$ formation on grains	& 0.707		& 1.00	\\
        \hline
      \end{tabular}
    \end{center}
  \end{table}
\end{center}

\begin{figure}[!htp]
\centerline{\includegraphics[height=175mm,clip=]{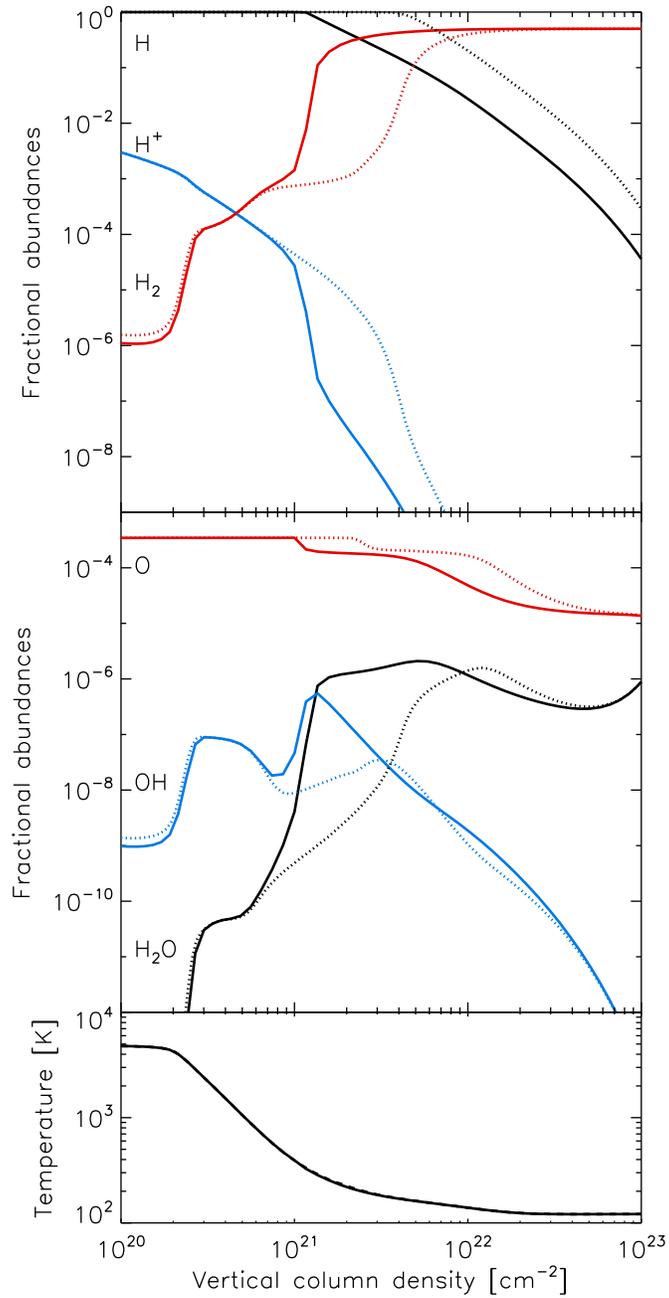}}
\caption{Hydrogen (top) and oxygen (middle) species at 1\,AU plotted 
vs.~vertical column density for old (dotted line) and new (solid line) 
three-body reaction rate coefficients discussed in GNI04 
(dotted line) and this paper (solid line). See Sec.~2.3 for details. 
With the new reaction rates, the H-$\hm$ transition, where 
$x({\rm H})= x({\rm H}_2)$, is shifted upwards by a factor of three 
in vertical column from 
$\Nh = 7\times 10^{21} \persqcm$ to $\Nh = 2\times 10^{21} \persqcm$.
The transition to water is also similarly shifted upwards. The solid 
curves represent model 1 in Tables 1 and 2.}
\end{figure}

% Sec. 2.4
\subsection{The Formation of $\hm$ on Dust Grains}

As discussed in Sec.~2.2, we previously considered only gas phase 
pathways for forming H$_2$ on the assumption that dust grains in 
disk atmospheres are too warm for efficient grain synthesis. This
idea is well founded in the standard theory of $\hm$ formation in 
interstellar clouds (e.g., Hollenbach and Salpeter 1971) and 
substantiated by low-temperature laboratory experiments using 
interstellar dust analogs and interpreted with theoretical models 
based on the physical adsorption of H atoms (e.g., Vidal et al.~2004; 
Perets et al.~2005, 2007). The efficiency of $\hm$ formation is close 
to unity in a narrow range (width $\sim 5$\,K) of temperature near 15\,K, 
depending on the experimental sample. 
At lower temperatures, incident H atoms are insufficently 
mobile, and at higher temperatures they leave the surface before 
forming molecules. Invoking a model based on chemical adsorption,
Cazaux and Tielens find that H$_2$ can form with moderate 
efficiency ($ \lesssim 0.2$) on warm dust with temperatures up to 
$~900$\,K (Cazaux \& Tielens 2004; Cazaux et al.~2005). We adopt
this model here because it is a reasonable extension of existing 
understanding and because it supported by laboratory experiments 
with graphite at intermediate temperature of a few hundred Kelvins 
(e.g., Zecho et al.~2004; Hornakaer et al.~2006a,b). 

We add grain formation to the program described earlier 
by writing the $\hm$ formation rate per unit volume in the 
standard form, i.e., 1/2 times the destruction rate of atomic H:
\be
R = \frac{1}{2}\,n(\h)v(\h)\, \nd <\pi a^2> \,\epsilon \, S(\Tg,\Td).
\ee
The two factors after the 1/2 give the incident H atom flux, the 
next two specify the grain surface area per unit volume, and the 
last two are the product of the formation efficiency and the sticking 
probability (or, more simply, the {\it overall} efficiency). We use 
GNI04 for the mean grain area per unit volume, based on the MRN 
distribution and an internal dust grain density of 3 $\gpercc$, 
to obtain (in units of $\ccps$),
\be
\label{grainformation1}
R = 4.32 \times 10^{-19}  \Tg^{1/2} \, n(\h)\nh \, \epsilon S
	\times (\frac{\rhod / \rhog}{0.01}) \, (\frac{\micron}{a_g}).
\ee
We use the right panel of Figure 2 of Cazaux et al.~(2005) for the 
formation efficiency $\epsilon$ as a function of the dust temperature 
$T_d$ for the one case they considered where the gas and dust 
temperatures are unequal, $T_g = T_d + 500{\rm K}$. A similar gaseous 
temperature enhancement is expected in the region in which the molecular 
emission is likely to arise. We estimate $S=0.1$ from Burke \& Hollenbach 
(1983).

The upper panel of
Figure 2 gives the results for the hydrogen family at $R=1$\,AU. 
Three sets of abundance curves are plotted vs.~vertical 
column density for three levels of grain formation: 
no grain formation (dotted line);  
grain formation with $a_{\rm geom}= 7.07 \micron $ (dashed curve);  
grain formation with $a_{\rm geom}= 0.707 \micron $ (solid curve);
they correspond respectively to models 1, 3 and 2 in Table 1.   
Note that the usual interstellar value is $a_g = 0.035\,\micron$.
In accord with the discussion in Sec.~2.1, the two choices of 
$a_{\rm geom}$ correspond to reduction factors of 20 and 200 in 
the grain area per H nucleus compared to ISM values. If we define 
the location of the H to H$_2$ transition by the vertical column 
where $x({\rm H})= x({\rm H}_2)$, then the transition occurs higher 
in the atmosphere (at smaller vertical columns) as the grain 
area increases. For the case $a_{\rm geom}= 0.707 \micron $, the 
H-$\hm$ transition, where $x({\rm H})= x({\rm H}_2)$, is shifted upwards 
by a factor of four in vertical column from 
$\Nh = 2.4\times 10^{21} \persqcm$ to $\Nh = 6\times 10^{20} \persqcm$.
The bottom panel of Figure 2 plots the temperature for the three cases 
in the upper panel. The dotted and solid temperature curves are roughly 
the same because they have the same grain area, i.e., 
$a_{\rm geom}= 0.707 \micron $. For the dashed curve,  
$a_{\rm geom}= 7.07 \micron $, the temperature is raised above 
the dust temperature because the gas and dust become less-well coupled 
thermally as the grain surface area decreases. Comparing the upper 
with the lower panel of Figure 2, we see that the H-$\hm$ transition 
for the solid curves ($a_{\rm geom}= 0.707 \micron $) occurs roughly 
in the middle of the transition region and that this model yields a 
significant column of warm water ($T > 300$\,K),  
$N(\htwoo) \approx 2 \times 10^{17}\, \persqcm$.

\begin{figure}[!htp]
\centerline{\includegraphics[height=175mm,clip=]{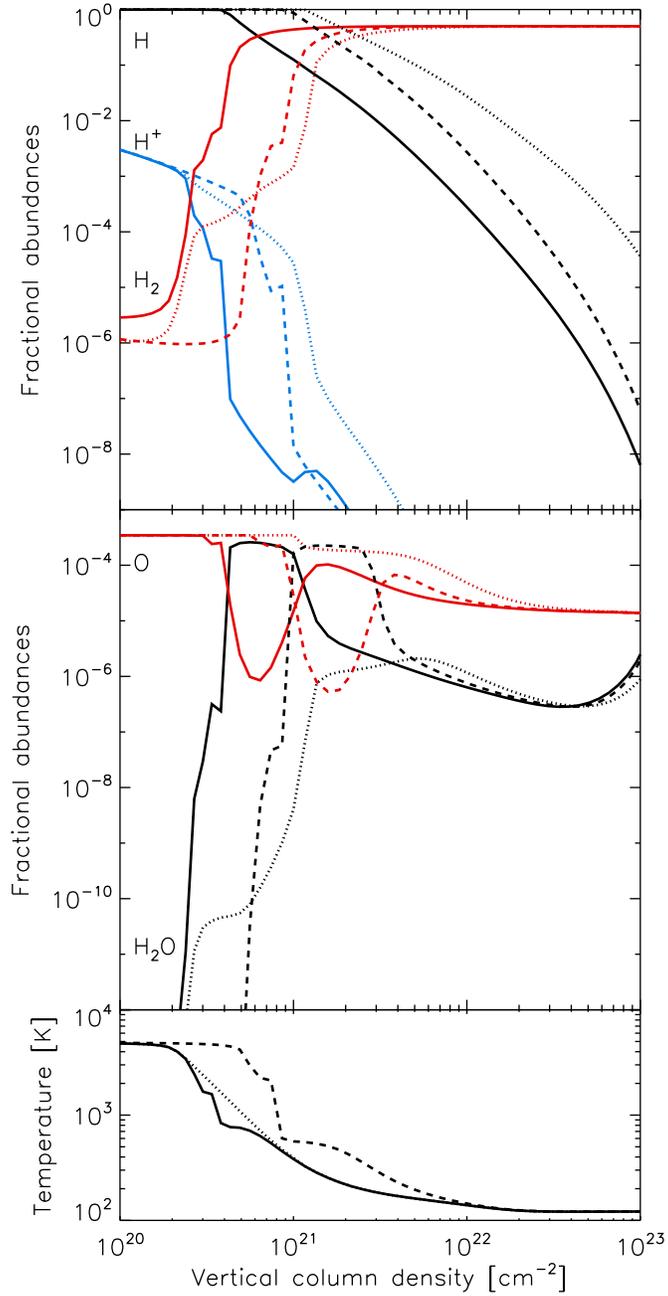}}
\caption{ Main hydrogenic species, atomic oxygen and water, and 
  gas temperature at 1\,AU
  plotted vs.~vertical column density for three cases of grain
  formation: no grain formation (dotted line); grain formation with
  $a_{\rm geom}= 7.07 \micron $ (dashed curve); grain formation with
  $a_{\rm geom}= 0.707 \micron $ (solid curve).  These three cases
  correspond respectively to models 1, 3, and 2 in Tables 1 and 2. }
\end{figure}

Figure 2 shows that grain formation of H$_2$ is important in the warm
transition region of the disk atmosphere. Increasing the abundance of
$\hm$ also increases the abundances of OH and $\htwoo$, which leads to
a decrease in the abundance of $\hp$ through the reactions in
Eqs.~\ref{ohionic} and~\ref{h2oionic}. This adds a degree of runaway
to the abundances since decreasing $\hp$ increases the amounts of OH
and $\htwoo$. The two main oxygen species, O and H$_2$O, are plotted
in the middle panel of Figure 2. The increased abundance of H$_2$
manifests itself in a significant enhancement of H$_2$O in the warm
transition region. Without grain formation of $\hm$, {\it full}
conversion of O to $\htwoo$ does not occur until a vertical column
$\Nh = 1.5 \times 10^{23}\, \persqcm$ has been reached, at which depth
dust cooling has overpowered gas heating and the gas and dust
temperatures are the same. At $R=1$\,AU, this temperature is of order
100\,K, so such a model cannot account for the warm water observed at
near-and MIR wavelengths. However, with grain formation of
$\hm$, full conversion occurs higher in the atmosphere where the gas
temperature is elevated above that of the dust.  For example, for the
case $a_{\rm geom}= 7.07 \micron $, the O-$\htwoo$ transition occurs
near $\Nh = 10^{21}\, \persqcm$ and, for $a_{\rm geom}= 0.707 \micron
$, it occurs at $\Nh = 4 \times 10^{20}\, \persqcm$.  In both cases,
there is a warm layer where almost of all the available oxygen is in
water (recall that $x_{\rm O}=3.5\times 10^{-4}$ and $x_{\rm
C}=1.4\times 10^{-4}$). For the case $a_{\rm geom}= 0.707 \micron $,
this layer contains a water column, $N(\htwoo) \approx 2 \times
10^{17}\, \persqcm$, and for $a_{\rm geom}= 7.07 \micron $, the water
column is $N(\htwoo) \approx 4 \times 10^{17}\, \persqcm$. This
increase occurs because the increase in $a_{\rm geom}$ implies a
decrease in dust cooling, which results in a shift downwards of the
warm region into denser gas.  These columns of warm water approach
the values obtained from recent {\it Spitzer} observations assuming
thermal populations, $\gtrsim 10^{17} \persqcm$ (Carr \& Najita
2008; Salyk et al.~2008). The OH abundance also rises as the 
transition to water is approached, reaching abundance levels 
$\sim 10^{-7}-10^{-6}$ and then decreasing rapidly with increasing 
vertical column of hydrogen.

Below these layers of near maximum water abundance high in the
atmosphere, the water abundance drops two orders of magnitude before
recovering at very large depths. The origin of this abundance swing
can be found in the approximate formula Eq.~\ref{closed_formula}.
Once $\hm$ has been fully formed, the $\htwoo / {\rm O}$ ratio is
governed by the relative importance of the temperature decline, as the
gas and dust temperatures become more and more closely coupled, and by
the decrease in the abundances of the ions $\hp$ and $\hep$ which
destroy $\htwoo$. With increasing depth, the decrease in temperature
first reduces the production of water, but then the attenuated X-ray
flux (and consequent low ion abundance) removes any effective
destruction so that its abundance increases again.

The full conversion to water is closely associated with the two other
chemical transitions, H to $\hm$ and C to CO. These transitions are
displayed in Figure 3 at two smaller radii, $R=0.25$ and $R=0.5$\,AU,
as well as at $R=1.0$\,AU (used in the previous figures). In all
cases, the transitions occur in sequence at increasing vertical
columns: first H to $\hm$ at the low abundance of $\hm$ ($\sim 10^{-6}$) 
needed to start the warm radical reaction sequence described in
Sec. 2.1; then C to CO, next O to $\htwoo$, and finally full conversion 
of H to $\hm$.  Full conversion of the residual O to $\htwoo$ occurs
where $\xhm \sim 0.1$.

\begin{figure}[!htp]
\centerline{\includegraphics[height=225mm,clip=]{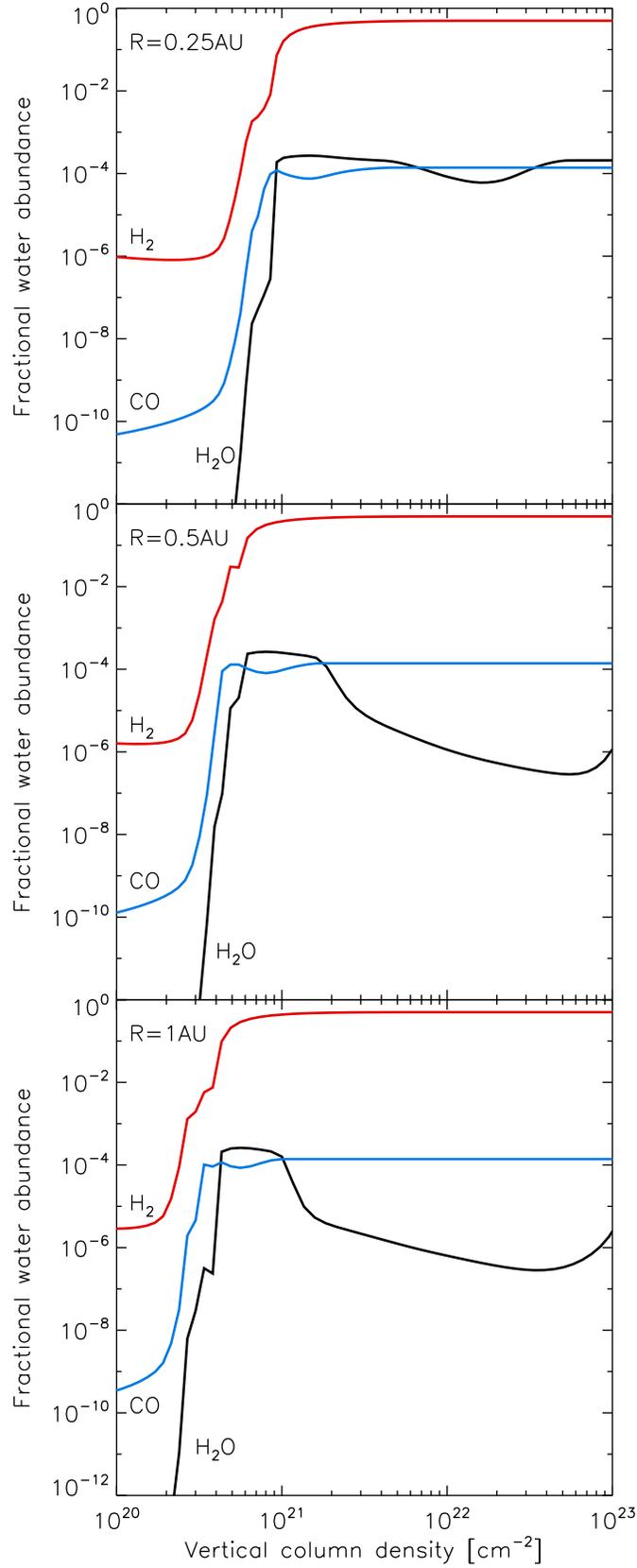}}
\caption{H$_2$, CO and water abundances at R=0.25 (top), 0.5 (middle)
  and 1.0 (bottom) AU for model 2 with H$_2$ formation on grains and 
  $a_{geom}=0.707\mu$m.}
\end{figure}

We conclude that the formation of $\hm$ on dust grains can play a 
key role in the synthesis of water in protoplanetary disk atmospheres. 
The efficiency of grain formation is determined overall by the effective 
surface area of the grains, which we have expressed in terms of the 
geometric mean $a_{\rm geom}$ of the minimum and maximum grain sizes in a 
MRN distribution.  Both $a_{\rm geom}$ and the efficiency of grain 
formation are affected by the amount of grain growth and settling.  
Figure 2 gives a concrete illustration of this effect at 1\,AU.
The top panel of shows that increasing $a_{\rm geom}$ from $0.707 \micron $ 
to $7.07 \micron$ shifts the H-$\hm$ transition downwards (to larger values 
of $\Nh$). This means that the layer where the water abundance is a maximum 
is also shifted downwards, as shown in the dashed curves in the middle panel 
of Figure 2. 

We have focused on two aspects of grain formation of $\hm$ that are 
important in disks, the extension of the standard theory to warm dust 
temperatures and the effect of grain growth and settling on the effective 
grain surface area. For the latter we use a single average parameter, 
the geometric mean of the minimum and maximum grain size ($a_{\rm geom}$), 
ignoring many complexities such as the nature of the grain surface and 
the variation of dust properties with grain size. We also assume that 
the same grain area applies to both the formation of $\hm$ and to the 
thermal coupling of the gas and the dust.  
The issue of $\hm$ formation on warm grains arose earlier in connection 
with ISO measurements of the pure rotational transitions of $\hm$ in photon 
dominated regions (PDRs) that seemed to require higher temperatures or 
more $\hm$ than predicted by PDR models (e.g., van Dishoeck 2004).   
Habart et al.~(2004) confronted this problem by considering the $\hm$ 
grain formation rate as a variable in fitting measured 0-0 S(3) to 
1-0 S(1) line ratios with a PDR model. They found ``normal'' (diffuse 
cloud) rate coefficients for high-excitation regions like the Orion bar 
and values enhanced by a factor of five for moderate excitation PDRs. 
They also determined a rough scaling between the $\hm$ 1-0 S(1) line 
and the strength of PAH features that suggested $\hm$ formation on PAHS 
might be responsible for the empirically determined formation rates. 
Habart et al.~(2004) and Cazaux et al.~(2006) showed that ``indirect'' 
chemisorption, where a second H atom reaches a chemisorption site 
from a physisorption site, could explain the empirical grain formation 
rates deduced for PDRs if the barrier between the two kinds of sites 
is small enough ($\sim 0.05$\,eV) and if a significant population of
small grains is present. While we do not consider $\hm$ formation 
by PAHs here because there are likely to be few PAHs in the inner regions 
of protoplanetary disks (see the Appendix), we do employ a reduced 
population of small grains for the synthesis of $\hm$. 

The question of whether a barrier as small a 0.05\.eV between physisorption
and chemisorption sites is appropriate may have been resolved by laboratory experiments with graphite at moderate temperatures which suggest that the 
formation of H-atom dimers (Hornak\ae r et al.~2006a) or clusters 
(Hornak\ae r et al.~2006b) may be an important step in grain formation of 
$\hm$. Although the Cazaux-Tielens theory of grain formation of $\hm$ 
yields a consistent picture for PDRs, it may not carry over to 
protoplanetary disk atmospheres of disks. The densities are much higher 
and the abundance of small grains is much less than in PDRs. However, 
the demonstrated sensitivity of molecular abundances to the abundance 
of $\hm$ may eventually lead to a better understanding of $\hm$ formation 
in disks when more extensive observations of species like water become 
available.  

% Sec. 2.5
\subsection{The Formation of Water on Dust Grains}

Water may also be formed by the sticking of O atoms and OH radicals 
on warm grains and subsequent reactions with adsorbed hydrogen atoms 
or other species. In regions where water is not already the dominant
oxygen species, atomic O is much more abundant than OH, and we consider
only this case. For this purpose, we adapt the formula for $\hm$ formation 
on grains (Eq.~\ref{grainformation1}) by dropping the 1/2 factor, noticing 
that $v({\rm O}) = (1/4) v(\h)$ and introducing a grain formation efficiency 
factor $\epsilon({\rm O})$ and a sticking factor $S({\rm O})$, 
\bdis
R(\htwoo)  =  2 \times 10^{-19} \ccps \Tg^{1/2} n({\rm O}) \nh
\edis
\be
\label{grainformation2}
\hspace{1.0in}\times (\frac{\rhod / \rhog}{0.01}) \,(\frac{\micron}{a_g}) \,
            \epsilon({\rm O})\,S({\rm O})
\ee           
We have implemented Eq.~\ref{grainformation2} using $\epsilon({\rm O})=1$ 
and $S({\rm O})=1$ to estimate the {\it maximum} role of water formation 
on grains. Figure 4 displays the changes in O and $\htwoo$ for the case 
of no grain formation, i.e., 
pure gas phase synthesis (dotted curve), grain formation of only $\hm$ 
with $a_{\rm geom}= 7.07 \micron$ (dashed curve), and grain formation 
of both $\hm$ and $\htwoo$ with $a_{\rm geom}= 0.707 \micron$ (solid curve). 
These cases correspond respectively to models 1, 3, and 4 in Tables 1 and 2. 
This model of maximal grain formation leads to two effects for $R = 1$\,AU. 
At very small columns,  $\Nh < 2 \times 10^{20}\, \persqcm$, the water 
abundance is increased by three dex to the $10^{-10}$ level, but this 
level of hot water is probably not observable. Grain formation of water 
also plays some role deeper down ($\Nh > 10^{21}\, \persqcm$) where gas 
phase synthesis is shut off by the low temperature. 
For $\Nh > 2 \times 10^{21}\, \persqcm$ at 1\,AU, the increased abundance 
of water should hardly affect the amount of observable warm water since 
it occurs where the temperature is close to 100\,K. 
However, inside of 0.5\,AU, water at large depths 
$\Nh > 2\times 10^{21}\, \persqcm$ will be observable because the gas 
is much warmer there. Nonetheless, the top layer of high-abundance 
water is the most important, and we conclude that grain formation of 
water is unlikely to play a dominant role in the inner disk.

\begin{figure}[!htp]
\centerline{\includegraphics[height=109mm,clip=]{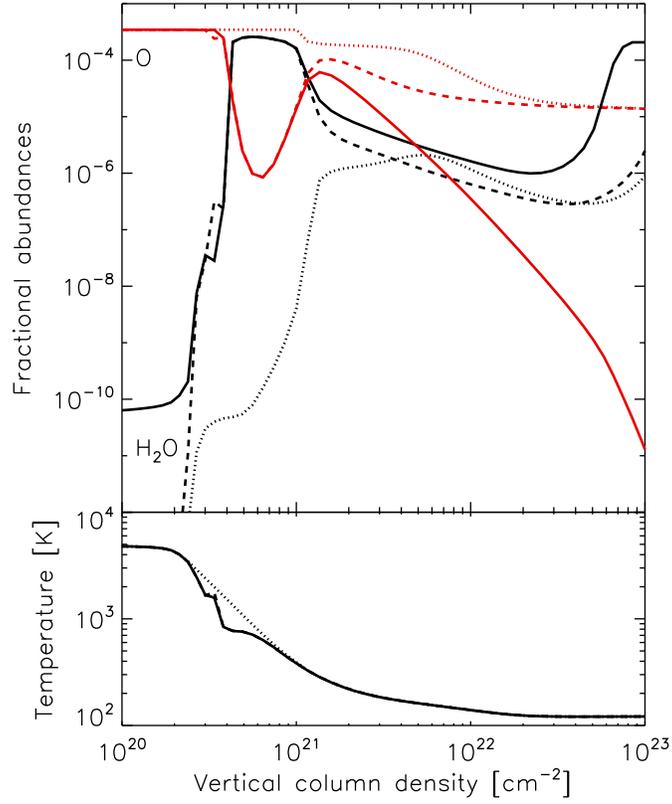}}
\caption{Oxygen species and temperature at 1\,AU vs.~vertical column 
density for three cases of $\hm$ and $\htwoo$ formation on grains: 
no grain formation (dotted line); $\hm$ formation on grains with 
$a_{\rm geom}= 0.707 \micron$ (dashed curve); - 
$\hm$ and $\htwoo$ formation on grains with 
$a_{\rm geom}= 0.707 \micron $ (solid curve). These three cases correspond 
respectively to models 1, 3, and 4 in Tables 1 and 2.}
\end{figure}
Another process that might contribute to the formation of water in the 
dense inner regions of a protoplanetary disk is radiative association,
\be
\label{radassn}
{\rm O} + \hm \ra \htwoo + h\nu.  
\ee 
To the best of our knowledge, this reaction has not been discussed in the 
literature. To exceed the production of water by grain formation, the 
rate coefficient for radiative association would have to be larger 
than $10^{-17}\,\ccps$. Although this is a typical rate coefficient 
for the radiative association of neutral species, it may not apply to 
water, at least without detailed consideration of the relevant 
potential energy curves. The rate coefficient for the radiative association 
of OH and H has been calculated by Field, Adams and Smith (1980). It is  
a strongly decreasing function of temperature, large at low temperatures 
($\gtrsim 10^{-14}\,\ccps$) but small in warm regions 
($\lesssim 10^{-18}\,\ccps$ above 300\,K). This process might be 
relevant for cool atomic H regions, but not in the atmospheres of the 
inner disks under consideration here. On the other hand, if the rate 
coefficient for Eq.~\ref{radassn} is significantly larger than 
$10^{-17}\,\ccps$ at moderate temperatures, this reaction would 
be an important source of water. 

% Sec. 2.6
\subsection{Mechanical Heating}

Finally we consider non-radiative heating processes that can affect the 
abundance and location of water. To illustrate this possibility, we adopt 
the mechanical heating model used in GNI04,
\be
\label{mechheat}
\Gamma_{\rm acc} = \frac{9}{4} \alphah \rhog c^2 \Omega,
\ee	
where $\rhog$ is the local gas mass density, $c$ is the isothermal sound 
speed, $\Omega$ is the angular rotation speed, and $\alphah$ is a  
phenomenological constant. GNI04 used values in the range 
$\alphah= 0.01-1.0$; they showed that the value 0.01 differs 
negligibly from the case $\alphah=0$ for pure X-ray heating. 
Mechanical heating might arise from viscous 
accretion generated by the MRI or from the interaction of a stellar 
wind with the upper layers of the disk. The change from the dotted to 
the dashed curves in Figure 5 shows the effect of increasing $\alphah$ 
from 0.01 to 1.0 for the case of no formation of $\hm$ on grains. 
As discussed in GNI04, mechanical heating causes the transition from 
high to low temperatures to start deeper down in the disk, near 
$\Nh = 10^{21}\, \persqcm$. It also produces a region of warm water 
($T > 300$\,K) at high abundance. The solid curve in Figure 5 shows that 
the combination of mechanical heating and grain formation of $\hm$ can increase 
the thickness of warm water before gas phase production is cut off at low 
temperatures. The column of warm water for model 6 is now 
$N(\htwoo) = 6.5 \times10^{17}\, \persqcm$ at 1\,AU, and even larger at 
smaller radii. These columns are of the same order as reported by Carr 
and Najita (2008) and Salyk et al.~(2008).  This result illustrates how 
mechanical heating can be important in determining the abundance of water  
and highlight the need for further studies of mechanical heating processes.

\begin{figure}[!htp]
\centerline{\includegraphics[height=109mm,clip=]{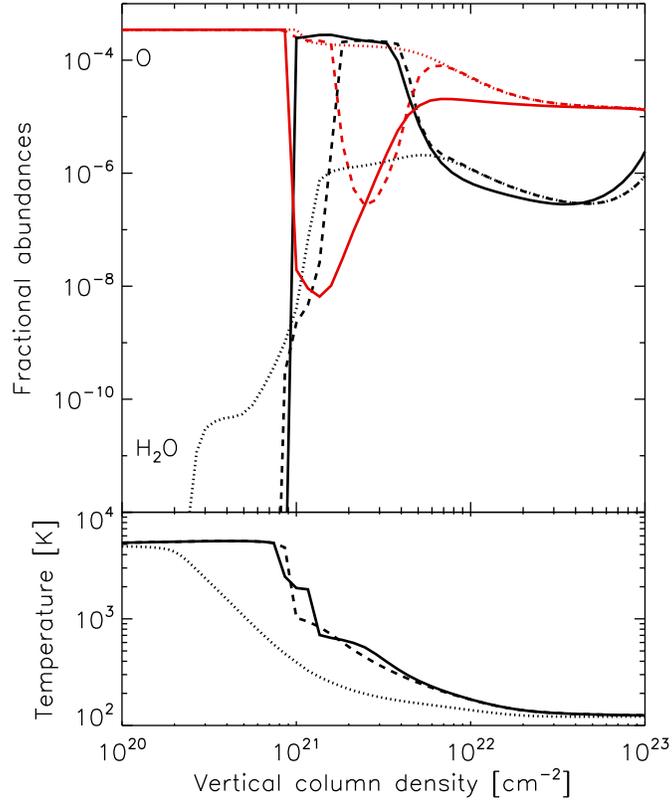}}
\caption{ Effect of mechanical heating on the abundance of
molecular hydrogen and water: 
$\alphah = 0.01$ and no grain formation (dotted curve); 
$\alphah = 1.0$ and no grain formation (dashed curve); 
$\alphah = 1.0$ with grain formation of $\hm$ and
$a_{\rm geom}= 0.707 \micron $ (solid curve).
These three cases correspond respectively to models 1, 5, and 6 
in Tables 1 and 2. 
}
\end{figure}

\section{Discussion}

In the previous section, we addressed the question of what 
determines observable levels of water in the inner region of a protoplanetary 
disk. One critical factor is the synthesis of $\hm$, which is required for 
the production of OH and $\htwoo$ from atomic O via neutral radical 
reactions. The gas must be warm in 
order for gas-phase synthesis to proceed, and the synthesis of $\hm$ must 
occur at moderate depths into the transition region in order for the $\hp$ 
abundance to be reduced by the attenuation of the X-rays and by the increase 
in density. All of these factors, the $\hm$ abundance, the temperature, and 
the level of ionization, can come together to produce maximum abundances of 
water in the thermal-chemical transition region that lies between the hot 
upper layer and the mid-plane of the atmosphere of an X-ray irradiated 
protoplanetary disk. 

We arrived at this conclusion by showing how sensitive the amount of warm 
water in protoplanetary disk atmospheres is to poorly known processes 
involving the $\hm$ molecule. We used the device of comparing the $\hm$ 
and $\htwoo$ abundances in the original model of GNI04 (corrected for the
error in the rate coefficient $k_5$ in Eq.~\ref{slowestneutral}) with the 
abundances calculated for new rate coefficients and processes. These 
include (1) the adoption of the  Flower and Harris (2007) rate 
coefficients for the three-body reactions Eq.~\ref{3bodyreacs}, (2) the 
introduction of grain formation of $\hm$ on the basis of the theory of 
Cazaux and Tielens (2003, 2004), Eq.~\ref{grainformation1},
(3) the assumption that $\htwoo$ is formed on grains at the 
maximum rate, Eq.~\ref{grainformation2}, and (4) mechanical 
heating of the atmosphere, Eq.~\ref{mechheat}. These changes were 
illustrated by the abundance patterns in the previous figures.
Although we emphasized water, the changes we have discussed also 
affect atomic O and OH and the species with which they interact. 

\begin{figure}[!htp]
\centerline{\includegraphics[height=109mm,clip=]{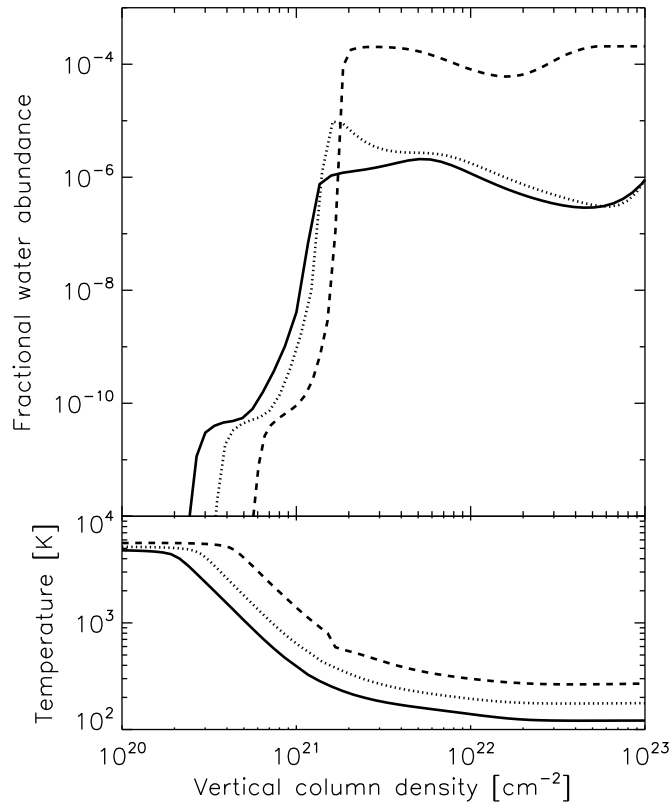}}
\caption{Water abundance and temperature profiles at R=0.25 (dashed),
  0.5 (dotted) and 1.0 (solid) AU for model 1 for the case of only 
  gas phase reactions (model 1) and $a_{geom}=0.707\mu$m.}
\end{figure}

An important measure of the changes is the column density of warm water, 
defined to have $T > 300$~K {\it and} an abundance large enough to 
potentially make a significant contribution to observable line fluxes. 
In practice, this column begins where the water abundance first achieves 
its maximum value and ends where the temperature drops below about 300\,K. 
These columns are given in Table 2 for the models considered 
in the previous section, along with the columns of warm OH and CO. The 
table includes the corresponding ranges in hydrogen column density ($\Nh$), 
temperature ($T$), electron fraction  ($x_{\rm e}$), and molecular hydrogen 
abundance ($\xhm$). Not only do these abundances express the physical 
conditions where the warm water occurs, but they also determine which 
collision partners might be important in exciting observable transitions. 
Electronic excitation of the low-lying rotational and ro-vibrational levels 
of water by electrons are typically $10^4$ times stronger than for $\hm$ 
(Faure and Josselin, 2008). Thus electronic excitation will not dominate 
in these warm water regions. But the H-$\hm$ transition may not be complete, 
so that atomic H collisional excitation is likely to play a role, not 
just $\hm$ collisions.  

\begin{center}
\begin{table*}
   \caption{Warm molecular columns}
   \begin{center}
     \begin{footnotesize}
     \begin{tabular}{lccccccc}  \hline \hline
    	\multicolumn{8}{c}{R=0.25~AU} \\
       \hline
       Model &  $N_{\rm H}$ [cm$^{-2}$] &  $T$[K]   & $x$(e)           
& $x$(H$_2$)      & $N$(H$_2$O) [cm$^{-2}$] & $N$(CO) [cm$^{-2}$] & $N$ 
(OH) [cm$^{-2}$] \\
       \hline
       1$^a$ & 1.7(21)-9.5(21)         & 300-560  & 2.0(-6)-4.2(-6)  &  
5.5(-2)-4.5(-1) & 1.1(18)                & 1.1(18)            &  
2.8(15) \\
       2$^b$ & 9.3(20)-9.5(21)         & 300-810  & 1.6(-6)-7.9(-6)  &  
7.3(-2)-4.9(-1) & 1.5(18)                & 1.1(18)            &  
2.4(15) \\
       3$^c$ & 1.7(21)-1.4(22)         & 310-650  & 1.5(-6)-4.2(-6)  &  
4.1(-2)-4.9(-1) & 2.8(18)                & 1.6(18)            &  
1.3(15) \\
       4$^d$ & 9.3(20)-9.5(21)         & 300-810  & 1.5(-6)-7.9(-6)  &  
7.5(-2)-4.9(-1) & 1.6(18)                & 1.1(18)            &  
2.4(15) \\
       5$^e$ & 2.1(21)-1.9(22)         & 300-720  & 7.4(-7)-4.1(-6)  &  
4.7(-2)-4.9(-1) & 3.7(18)                & 2.3(18)            &  
1.4(15) \\
       6$^f$ & 1.4(21)-1.9(22)         & 300-1000 & 5.7(-7)-1.1(-5)  &  
3.4(-1)-4.9(-1) & 4.0(18)                & 2.3(18)            &  
7.9(14) \\
       \hline
       \multicolumn{8}{c}{R=0.5~AU} \\
       \hline
       Model &  $N_{\rm H}$ [cm$^{-2}$] &  $T$[K]   & $x$(e)           
& $x$(H$_2$)      & $N$(H$_2$O) [cm$^{-2}$] & $N$(CO) [cm$^{-2}$] & $N$ 
(OH) [cm$^{-2}$] \\
       \hline
       1$^a$ & 1.2(21)-2.1(21)        & 320-510   & 4.7(-6)-5.7(-6)  &  
4.4(-2)-2.6(-1) & 6.2(15)                & 1.0(17)            &  
7.3(14) \\
       2$^b$ & 4.3(20)-2.1(21)        & 320-790   & 5.1(-6)-9.2(-6)  &  
2.9(-2)-4.7(-1) & 3.0(17)                & 2.0(17)            &  
1.7(15) \\
       3$^c$ & 9.2(20)-4.8(21)        & 330-600   & 2.0(-6)-3.5(-6)  &  
3.5(-2)-4.7(-1) & 7.9(17)                & 4.9(17)            &  
1.0(15) \\
       4$^d$ & 5.5(20)-2.1(21)        & 320-790   & 5.0(-6)-8.5(-6)  &  
3.1(-2)-4.7(-1) & 3.1(17)                & 2.0(17)            &  
1.7(15) \\
       5$^e$ & 1.9(21)-6.4(21)        & 300-670   & 1.6(-6)-2.8(-6)  &  
6.6(-2)-4.4(-1) & 9.2(17)                & 6.5(17)            &  
8.8(14) \\
       6$^f$ & 6.2(20)-5.5(21)        & 320-1090  & 1.5(-6)-1.1(-5)  &  
2.0(-1)-5.0(-1) & 1.1(18)                & 5.6(17)            &  
7.9(14) \\
       \hline
       \multicolumn{8}{c}{R=1~AU} \\
       \hline
       Model &  $N_{\rm H}$ [cm$^{-2}$] &  $T$[K]   & $x$(e)           
& $x$(H$_2$)     & $N$(H$_2$O) [cm$^{-2}$] & $N$(CO) [cm$^{-2}$] & $N$ 
(OH) [cm$^{-2}$] \\
       \hline
       1$^a$ & $>$ 1.2(21)             & $>$300   & $>$1.2(-5)      &  
$<$7.5(-3)      & 4.4(12)                & 6.6(15)            &  
6.2(13) \\
       2$^b$ & 4.3(20)-1.2(21)         & 330-770  & 6.3(-6)-8.1(-6) &  
9.7(-2)-4.6(-1) & 1.6(17)                & 9.9(16)            &  
7.9(14) \\
       3$^c$ & 1.0(21)-2.9(21)         & 310-560  & 2.7(-6)-3.3(-6) &  
6.4(-2)-4.5(-1) & 4.0(17)                & 2.9(17)            &  
7.9(14) \\
       4$^d$ & 4.3(20)-1.2(21)         & 330-770  & 6.2(-6)-8.1(-6) &  
9.9(-2)-4.6(-1) & 1.6(17)                & 9.2(16)            &  
6.6(14) \\
       5$^e$ & 1.8(21)-3.8(21)         & 340-630  & 1.6(-6)-2.1(-6) &  
7.4(-2)-3.7(-1) & 4.5(17)                & 3.7(17)            &  
5.1(14) \\
       6$^f$ & 1.0(21)-3.8(21)         & 300-1020 & 1.6(-6)-8.0(-6) &  
4.6(-1)-5.0(-1) & 6.5(17)                & 3.3(17)            &  
1.6(14) \\
       \hline
       \multicolumn{8}{c}{R=2~AU} \\
	\hline
       Model &  $N_{\rm H}$ [cm$^{-2}$] &  $T$[K]   & $x$(e)           
& $x$(H$_2$)     & $N$(H$_2$O) [cm$^{-2}$] & $N$(CO) [cm$^{-2}$] & $N$ 
(OH) [cm$^{-2}$] \\
       \hline
       1$^a$ & $>$ 9.4(20)            & $>$300   & $>$3.7(-5)       &  
$<$1.1(-3)      & 3.0(11)                & 5.0(13)            &  
3.2(13) \\
       2$^b$ & 3.9(20)-9.4(20)        & 300-720  & 6.6(-6)-8.6(-6)  &  
1.4(-1)-4.4(-1) & 1.0(17)                & 8.4(16)            &  
1.1(15) \\
       3$^c$ & 9.4(20)-1.8(21)        & 310-500  & 3.4(-6)-3.7(-6)  &  
1.0(-1)-4.1(-1) & 1.7(17)                & 1.7(17)            &  
6.2(14) \\
       4$^d$ & 3.9(20)-9.4(20)        & 300-720  & 6.6(-6)-8.6(-6)  &  
1.4(-1)-4.5(-1) & 1.0(17)                & 8.4(16)            &  
1.2(15) \\
       5$^e$ & 1.8(21)-3.4(21)        & 320-600  & 1.4(-6)-1.9(-6)  &  
9.8(-3)-3.4(-1) & 3.0(17)                & 3.2(16)            &  
6.2(14) \\
       6$^f$ & 4.5(20)-2.9(21)        & 310-1170 & 1.7(-6)-1.1(-5)  &  
3.0(-1)-5.0(-1) & 5.7(17)                & 2.8(17)            &  
4.4(14) \\
       \hline
       \multicolumn{8}{l}{$^a$Gas phase only, $a_{geom}=0.707\mu$m, $ 
\alpha=0.01$} \\
       \multicolumn{8}{l}{$^b$H$_2$ formation on grains,  
$a_{geom}=0.707\mu$m, $\alpha=0.01$} \\
       \multicolumn{8}{l}{$^c$H$_2$ formation on grains,  
$a_{geom}=7.07\mu$m, $\alpha=0.01$} \\
       \multicolumn{8}{l}{$^d$H$_2$ and H$_2$O formation on grains,  
$a_{geom}=0.707\mu$m, $\alpha=0.01$} \\
       \multicolumn{8}{l}{$^e$Gas phase only, $a_{geom}=0.707\mu$m, $ 
\alpha=1.0$} \\
       \multicolumn{8}{l}{$^f$H$_2$ formation on grains,  
$a_{geom}=0.707\mu$m, $\alpha=1.0$} \\
   \end{tabular}
   \end{footnotesize}
\end{center}
\end{table*}
\end{center}

Table 2 contains information about the radial variation of the warm
molecular columns of $\htwoo$, OH and CO. The warm water column
$N(\htwoo)$ rapidly increases with decreasing radius, faster than
$1/R$ for $R \lesssim 0.5$\,AU. The inner disk gets warmer due 
to increased accretion heating, which reduces the cooling of the gas 
by the dust. All of the models yield $N(\htwoo) > 10^{18} \persqcm$ 
for small radii ($R \lesssim 0.5$\,AU). This result
opens up the possibility that the water lines may have broad wings for
inclined disks. Detailed excitation calculations are needed to deal
with this issue, not just for water but for other molecules for which 
line shape measurements are feasible, especially CO. 
Accretion heating becomes increasingly 
important with decreasing $R$, to the point
where pure gas phase processes (model 5 in Table 1) give about the
same water columns as model 6 (with $\hm$ grain formation). This can
be seen in detail in Figure 6 which shows water abundance profiles at
several radii for pure gas phase chemistry.

The results of the few reported observations of water in disks around 
young low-mass stars (Carr et al.~2004; Carr and Najita 2008; Salyk 
et al.~(2008) have been expressed in terms of a column density and an 
excitation temperature. For the typical T Tauri star 
AA Tau, Carr and Najita find  $N(\htwoo) = 6.5 \times 10^{17} \persqcm$ 
and a characteristic excitation temperature of $\sim 575$\,K. The same 
quantities for OH and CO are $N(\oh) = 8.1 \times 10^{16} \persqcm$  
(characteristic temperature of $\sim 525$\,K) and 
$N(\co) = 4.9 \times 10^{17} \persqcm$ 
(characteristic temperature of $\sim 900$\,K).  Salyk et al.~(2008) report
similar water columns for two other T Tauri stars. These column densities
have been derived assuming thermal populations, so they may be 
underestimates or overestimates, depending on the excitation conditions. 
Although the model calculations in this paper were done primarily to 
illustrate how certain physical and chemical processes affect the 
amount of observable water in protoplanetary disks -- and not to model
any particular disk, it is of some interest to compare the observations 
with the results in Table 2. The presumed improvements in the treatment 
of the formation of $\hm$, especially grain formation, bring the 
calculated column of warm water close to the observed range 
$N(\htwoo) > 10^{17}-10^{18} \persqcm$. If the heating of the atmosphere
is increased, the reported columns $N(\htwoo) = 6-8 \times 10^{17} \persqcm$
can be reached or even exceeded. In drawing this conclusion, we have not 
varied the density distribution or the X-ray irradiation, which can affect 
the water column. For example, the X-ray emission of AA Tau appears to be 
typical for an active classical T Tauri star (Grosso et al.~2007), with a 
minimum or quiescent X-ray luminosity $\Lx = 5 \times 10^{30} \ergps$ and 
temperature $\Tx = 2.5$\,keV. Both parameters are larger than used 
here based on the 
GNI model, $\Lx = 2 \times 10^{30} \ergps$ and temperature $\Tx = 1.0$\,keV. 
In addition to the effects of choices in model parameters, the predictions
in Table 2 are subject to the limitations of our demonstration model (GNI04), 
e.g., the deficiencies in the thermal model and specifically the neglect
of water heating and cooling, as discussed in Sec.~2.1.

One feature of the model calculations is that the column density of 
OH is much less than that of water. According to the last column of 
Table 2, the typical column density of warm OH in the inner disk is, 
to within a factor of two, $N(\oh) \sim 10^{15} \persqcm$. The ratio 
of the warm $\htwoo$ to the warm OH column density ranges from a few 
hundred to several thousand, due mainly to the changes in the chemical 
model. The measured warm OH column for AA Tau is 
$N(\oh) = 8 \times 10^{16} \persqcm$, and none of the entries in Table 2 
approach this value. This points to a deficiency in our calculation of 
the OH abundance. Staying within the confines of gas-phase chemistry, 
this problem might be solved by including FUV photo-dissociation of OH 
and $\htwoo$, taking into account that the cross section for water 
is several times larger than for OH. Preferential photo-dissociation of 
water might then lead to a significant increase in the amount of OH. 
An extreme example of this effect is the detection in the L band 
(3$\micron$) of the lines of OH but not $\htwoo$ in the disks around 
two Herbig Ae stars (Mandell et al.~2008). In this case there is ample 
FUV from the A star to destroy circumstellar water in favor of OH. For 
T Tauri star disks, however, the stellar FUV flux is much smaller, and 
it may be unable to shift the balance from $\htwoo$ to OH, especially 
in the presence of neutral gas phase synthesis reactions which tend to drive 
OH to $\htwoo$. This is the situation in our model, despite the fact that 
the rate coefficient for the reaction that destroys water, Eq.~\ref{h2oionic}
($\hp + \htwoo \rightarrow \htwoop + \h$; Anicich 1993), is four 
times larger than that for OH, Eq.~\ref{ohionic}
($\hp + \oh \rightarrow \ohp+ + \h$; UMIST guess, Woodall et al.~2007).
However, Bergin et al.~(2003) have pointed out that the photo-dissociation 
of water is sensitive to the spectral shape of the stellar FUV, in 
particular to the strength of the Lyman-$\alpha$ line, which dominates 
the FUV emission in some T Tauri stars such as TW Hya. Thus sources 
with stronger Ly-$\alpha$ emission would be more effective in destroying
water. On the other hand,  OH can also be destroyed by Ly-$\alpha$ 
radiation (van Dishoeck and Dalgarno 1983). The rates of direct 
destruction of OH and $\htwoo$ by X-rays are about the same, since both 
cases depend largely on K-shell absorption by a single oxygen atom. 

Going beyond pure gas-phase chemistry, OH may be formed by desorption 
from icy grain mantles transported from cold regions to the warm source 
of the observed MIR lines within an AU or so from the star. The release 
could proceed by either thermal or photo-desorption (Ciesla and Cuzzi 
2006). The latter process has been suggested to be the origin of the 
highly excited OH ground rotational lines seen in HH 211 (Tappe et al.~2008). In this case, the FUV radiation is shock generated, and it can 
desorb both OH and $\htwoo$ (e.g., Andersson and van Dishoeck 2008). 
When the water is photo-dissociated, OH is formed in a highly excited 
state, and the excited OH rotational lines are produced by fluorescence. 
Similar processes may be at work in T Tauri disks, mediated by stellar 
FUV or X-ray radiation.    

A last comparison between the present model calculations and the 
limited observations to test them involves CO. According to Table 2, 
the model calculations predict comparable warm columns of $\htwoo$ 
and CO. The average ratio of water to CO to water columns in Table 2 
is 1.6, independent of radius, with small variations of order 25\%. 
Carr and Najita (2008) report a ratio of 3/4 for AA Tau, in rough accord 
with our model which ascribes most of the warm water and CO to the
thermal-chemical transition region. The higher temperature and smaller 
emitting area of the CO lines compared to the water lines in AA Tau 
suggest that the CO and the water probe somewhat different radial 
distances.  While the CO likely arises from small radii ($R < 1$\,AU), 
the water may arise from radii out to $\gtrsim 2$\,AU (Carr \& Najita 
2008). The individual column densities of warm CO and water reported 
in Table 2 are consistent with the properties of the molecular emission 
reported 
for AA Tau.
 
It is also of interest to compare our results with other chemical 
calculations of the inner disk, e.g., Markwick et al.~(2002), Ag\'{u}ndez 
et al.~(2008), Woods and Willacy (2008), and Gorti and Hollenbach (2008). 
These comparisons are characterized by the fact that disk chemistry 
models differ widely in their underlying assumptions as well as in 
execution. None of these studies include mechanical heating, and only 
Woods and Willacy consider the recent theory of grain formation of $\hm$ 
by Cazaux and Tielens (as do Nomura et al.~2009). For example, Markwick 
et al.~(2002) report only total molecular column densities with no 
information about the vertical abundance distribution. At 1\,AU, they 
obtain $N(\htwoo) = 1.6 \times 10^{22}\,\persqcm$ and 
$\Nh = 1.6 \times 10^{26}\,\persqcm$, for an average water abundance 
$\xhtwoo = 10^{-4}$. Since they include only viscous heating, their 
disk atmosphere is much cooler than obtained in models that include 
stellar irradiation, and most of their calculated water column will 
not be warm. Markwick et al.~do not discuss OH.

Ag\'{u}ndez et al.~(2008) developed a time-dependent model oriented towards the 
synthesis of complex molecules with neutral radical reactions. In Sec.~4 
of their paper, they use the D'Alessio (2001) model exposed to stellar and
interstellar FUV radiation, but without X-rays. They assume equal gas and 
dust temperatures and start with a molecular mixture with all oxygen in 
water that is not in CO. They obtain the following steady-state columns 
inside 3\,AU, approximately independent of radius:
$N(\co) = 10^{18}\,\persqcm$, 
$N(\htwoo) \approx 10^{17}\,\persqcm$ and 
$N(\oh) = 10^{14}\,\persqcm$. Although these values are similar to those 
calculated here, 
the molecules will not be as warm nor produce as much emission 
because the gas temperature is the same as the dust temperature. 

Woods \& Willacy (2008) have developed a thermal-chemical model based on 
the D'Alessio (2001) disk structure that includes stellar and interstellar 
FUV radiation and X-rays. Their main emphasis is on the carbon isotopes, 
but they also report on the total water column density and give contours 
of water and OH in the inner disk. They show that the regions of the disk 
atmosphere, where the average temperatures of the OH and water emission are
the same as found for these species in AA Tau (Carr and Najita 2008), have 
local abundances in the model that are similar to those reported by Carr 
and Najita. No quantitative information is provided on the column 
densities of warm OH and water.

Gorti and Hollenbach (2008) have calculated a wide variety of atomic and 
molecular lines with a full thermal-chemical model that is similar to but
more general than the one used here. It includes X-rays as well as 
stellar FUV, although the X-ray heating is significantly less than in our 
model. Gorti and Hollenbach also include cosmic rays, whereas we believe, 
by analogy with the much less powerful solar wind, that they are blown 
away from the inner protoplanetary disk by the stellar wind of the YSO
(Glassgold, Najita and Igea, 1997). A key element of their model is  
its use of a significant population of small grains and PAHS to heat the 
gas via the photoelectric effect. However, they do not include mechanical 
heating, which we find plays an important role in the inner disk. Gorti 
and Hollenbach also simultaneously solve the equation for hydrostatic 
equilibrium, so that their disk atmosphere is more flared than ours. The 
column density of water is smaller than calculated here, and the line 
strengths are modest because the warm water is at smaller densities due 
to disk flaring (D. Hollenbach, private communication 2009). Consequently, 
they predict that only a few MIR lines should be detectable by the 
{\it Spitzer} IRS, in contrast with the rich emission spectra reported 
by Carr and Najita (2008) and Salyk et al.~(2008). On the other hand, 
we do not treat hydrostatic equilibrium here, and our future modeling 
will need to include it in order to assess its importance for the 
formation of $\hm$ and the generation of significant columns of warm water.

\section{Conclusions}

Our results indicate that the conditions that lead to strong water
emission in the infrared, i.e., large column densities of warm water
molecules near the disk surface, can be produced {\it in situ} under 
certain conditions in an X-ray irradiated disk model without 
requiring radial or vertical transport of water to the disk surface.  
In the inner disk ($\lesssim 0.25$\,AU), abundant warm water 
can be produced through gas-phase processes and only X-ray heating.
Efficient $\hm$ formation on warm grains and 
non-radiative heating appear to be capable of enlarging the radial 
range over which warm water emission is produced.  These processes, 
when taken together, can account roughly for the column densities 
and emitting areas of the MIR water emission reported for T Tauri 
disks by Carr \& Najita (2008) and Salyk et al.~(2008).

We find that the column density (and temperature) of water at the
disk surface depends significantly on the grain surface area
in the disk atmosphere (i.e., on the extent of grain growth and
settling) and on the heating of the atmosphere (and therefore perhaps
on the disk accretion rate).  Because both the disk accretion rate and 
the extent of grain settling are believed to vary among T Tauri disks, 
our model results suggest that the strength and character of MIR 
water emission from disks may also show some diversity.  
In contrast, because much less restrictive conditions are needed to 
produce large column densities of warm water close in ($< 0.25$\,AU), 
water emission from this inner region of the disk may be quite common. 

More specifically, for the grain areas that we use, we find that 
sources with modest amounts of grain growth 
(e.g., a factor of 20 decrease in area compared to the ISM) can 
produce large column densities of warm water within 1\,AU.  
Even larger columns result for ten times more grain growth (see Table 2), and 
we might expect this trend to continue for a growth factor of 1000.
Recent studies of the MIR spectral energy distributions of T Tauri stars 
suggest that disks have experienced grain growth factors spanning a  
similar range (factors 10-1000; Furlan et al.~2006) as those 
considered in our models.  Thus, we might expect that T Tauri disks 
that show different amounts of grain growth may display a diversity in 
the strength and character of their water emission.

Similarly, T Tauri disks are also believed to have accretion rates 
that vary by at least two orders of magnitude (Hartmann et al.~2006).  
Stellar X-ray luminosities also vary significantly.
Since these factors influence the ionization and amount of surface heating, 
they may also contribute additional diversity to the water emission 
spectra of T Tauri disks. Large accretion rates can enhance the heating 
in the atmosphere and increase the strength of the water emission. 
A higher X-ray luminosity would produce both stronger ionization and 
heating. According to the simplified theory of Sec.\ 2.2, heating effects  
are more important than the ionization level. Such trends may be revealed in 
{\it Spitzer} IRS data sets that are being collected currently. However, 
if radial or vertical mixing is efficient, it might diminish or erase the 
correlations that would be predicted by our models. 

In addition to searching for correlations in large data sets, it may 
also be interesting to explore individual objects in greater detail 
in the context of our models. In the case of AA Tau, the analysis 
of Furlan et al.~(2006) suggests that the dust in the disk surface 
may be depleted by a factor of nearly 1000. Our calculations suggest
that the observed warm water columns can be accounted for, even for
this degree of depletion without non-radiative heating. Thus in
this case, mixing and transport appear to be unnecessary to account
for the observations. 

We have identified two major uncertainties in the chemistry of H$_2$
that bear on the amount of water and other species: (a) {\it
Three-body formation of $\hm$} - Although the use by Flower and Harris
of a new equilibrium constant represents an important step forward,
the data on which these reactions are based are old measurements of
the low-energy collisional excitation of $\hm$ by atomic hydrogen
(Jacobs et al.~1967; Baulch et al.~2005). New experiments are needed.
(b) {\it Grain formation of $\hm$} - The adopted rate coefficients are
based on theory (Cazaux and Tielens 2002, 2004).
Although this model is consistent with observations of $\hm$ lines 
in PDRs (Habart et al. 2004, Cazaux and Tielens 2006), its applicability
to the dense and dust-depleted atmospheres of protoplanetary disks is
not yet fully established.
 
Regarding non-radiative heating, we have used Eq.~\ref{mechheat} from 
GNI04, who suggested two ways of heating the transition region above 
the level achievable with X-rays, dissipation of turbulence and the 
interaction of the stellar wind with the disk atmosphere. A sound 
basis for these processes has not yet been laid, but progress should be 
possible with improved numerical simulations of the magneto-rotational 
instability (e.g., Turner and Sano 2008; Terquem 2008) and of the 
wind-disk interaction (e.g., Li and Shu 1996; Matsuyama et al.~2009).  
Our results highlight the need for further basic studies of the heating 
generated by the MRI and by the interaction of the wind and the disk.

Given these current uncertainties, and the potential sensitivity of water 
emission to multiple 
effects (X-ray irradiation, grain growth, disk accretion, transport, etc.), 
it may be difficult to diagnose the role of such important (but elusive) 
processes as radial and/or vertical transport from observations of water 
{alone.  It would be worthwhile to establish diagnostics for other 
relevant regions of a protoplanetary disk such as the disk midplane and 
the outer disk. An interesting example is provided by the discussion
by Woods and Willacy (2007) of the formation of benzene near the midplane 
of the inner part of a protoplanetary disk.}

\section{Acknowledgements} 

This work has been supported by NSF grant AST-0507423, NASA grants
NNG06GF88G and  1322305 to UC Berkeley. We would like to thank 
John Black, David Hollenbach and Xander Tielens for discussions 
of chemical processes in disks and David and Xander for helpful 
comments on our manuscript.

\appendix

\section{PAH Formation of $\hm$ in Protoplanetary Disks}

Duley and Williams (1993) suggested that the $\hm$ forms on PAHs in 
carbonaceous interstellar dust, and Bauschlicher (1998) demonstrated 
that it was energetically favorable for PAH cations to acquire an H 
atom and then form $\hm$ in a subsequent collision with atomic H. 
He suggested that this was a way to synthesize $\hm$ in the gas phase. 
Rauls and Hornek\ae r (2008) have recently discussed how super-hydrogenated 
neutral PAHs might also play a role in the gas-phase synthesis of $\hm$. 
Jonkheid et al.~(2006) included $\hm$ formation by PAHs in modeling the 
transitional disk around HD\,141569. Habart et al. (2004) adduced possible
evidence for $\hm$ formation by PAHs in PDRs.

As a basis for evaluating the possibility that PAH formation of $\hm$
is relevant in the molecular transition region  of disks, we make a crude
estimate of the rate coefficient  for the Bauschlicher process,
\be
\label{bausch}
({\rm PAH}^+ + \h) + \h \rightarrow {\rm PAH}^+ + \hm,
\ee
where the symbol (PAH$^+$ + H) represents a stable PAH cation with an
added H atom. The effective rate coefficient $R_{\rm PAH}$ for this process, 
which can then be compared with the rate coefficient $R$ for grain 
formation in Eq.~\ref{grainformation1}, can be written as,
\be
R_{\rm PAH} = k_{\rm PAH} \, f({\rm PAH}^+ + \h)\, x_{\rm PAH},
\ee   
where $k_{\rm PAH}$ is the {\it unknown} rate coefficient for
reaction~\ref{bausch}, $f({\rm PAH}^+ + \h)$ is the fraction of
appropriately hydrogenated PAH cations, and $x_{\rm PAH}$ is the total
gas phase abundance of PAHs in all ionization states.  We make a rough
guess that $k_{\rm PAH} \sim 10^{-10}\, \ccps$ and assume that $f({\rm
PAH}^+ + \h) \sim 10^{-2}$ and $x_{\rm PAH} \sim 10^{-7}$, an upper 
limit to the range deduced by Geers et al.~(2006) from {\it Spitzer}
observations of disks. The result,$R_{\rm PAH}  \sim 10^{-19}\, \ccps$, 
is only slightly smaller than given by Eq.~\ref{grainformation1} 
for grain formation of $\hm$.
	
Each of the factors in our estimate of $R_{\rm PAH}$ is uncertain by
at least one dex, so we cannot rule PAH formation of $\hm$ in or out
on this basis. However, we suspect that it is unimportant for the inner 
regions of protoplanetary disks because PAHS close to YSOs are likely 
to be destroyed by stellar X-rays. We estimated the total abundance 
$x_{\rm PAH} \sim 10^{-7}$ on the basis of the low detection frequency 
of PAH IR features in {\it Spitzer} observations of T Tauri stars 
(Geers et al.~2006; $\sim 10\, \% $) and Class I YSOs (Geers et al.~2008; 
$\lesssim 2 \%)$.  Even where observed, the PAHs may occur mainly at 
large radii, as for Herbig stars (Habart et al.~2006; Geers et al.~2007; 
Goto et al.~2008).  Laboratory experiments with the Synchrotron Radiation 
Facility in Grenoble show that PAHs and very small particles are destroyed 
by hard X-rays (Guegeon 1998; Mitchell et al.~2002; see also Sec.~6 of the
review by Glassgold, Feigelson \& Montmerle 2000). Indeed, Voit (1992)
invoked X-ray destruction of PAHs to explain the difference between
the observed PAH features seen in starburst galaxies and AGN. Finally,
Siebenmorgen and Kreugel (2008) have recently modeled the PAH
abundance in T Tauri disks and find that energetic photons destroy
PAHS inside of 10\,AU. It is for these reasons in support of a low 
PAH abundance that we tentatively ignore PAH formation of $\hm$ for 
the inner part of the atmospheres of protoplanetary disks.

\section{References} 

\noi Ag\'{u}ndez, M., Cernichara, J. \& Goicoechea, 2008, \aap, 483, 831 

\noi Andersson, S., Al-Halabi, A., Kroes, G.-J. \& van Dishoeck, D. F. 2008, 
	J. Chem. Phys., 124, 064715

\noi Anicich, V. 1993, J. Phys. Chem. Ref. Data, 22, 1469 

\noi Baulch et al.~1992, J. Phys. Chem. Ref. Data, 21, 411

\noi Baulch et al.~2005, J. Phys. Chem. Ref. Data, 34, 757

\noi Bauschlicher, C. W. 1998, \apj, 509, L125

\noi Bergin, E., Calvet, N., D'Alessio, P. \& Herczeg, G. J. 2003, 
\apj, 591, L159,

\noi Burke, J. R. \& Hollenbach, D. J. 1983, \apj, 265, 223

\noi Carr, J. S., Tokunaga, A. \& Najita, J. 2004, \apj, 603, 213

\noi Carr, J. S. \& Najita, J. 2008, Science, 319, 2008 

\noi Cazaux, S. \& Tielens 2002, A. G. G. M., \apj, 575, L29

\noi Cazaux, S. \& Tielens 2004, A. G. G. M., \apj, 604, 222

\noi Cazaux, S., Caselli, P., Tielens, A. G. G. M., Le Bourlet, J. \&
Walmsley, M. 2005, J Phys, Conf. Series, 6, 155

\noi Cazaux, S., Caselli, P., Walmsley, M. \& Tielens, A. G. G. M., 
2006 in Astrochemistry: Recent Successes and Current Challenges, 
eds. D. C. Lis, G. A. Blake \& E. Herbst, (Cambridge: Cambridge), p.~325

\noi Ciesla, F. J. \& Cuzzi J. N. 2006, Icarus 181, 178

\noi Cohen, N. \& Westberg, K. R. 1983, J. Phys. Chem. Ref. Data, 12, 531

\noi D'Alessio, P., Calvet, N., Hartmann,L., Lizano, S.
Cant\'{o}, J. 1999, \apj, 527, 893

\noi D'Alessio, P., Calvet, N. \& Hartmann, L. 2001, \apj, 553, 321

\noi Duley, W. W. \& Williams D. A. 1993, \mnras, 260, 37

\noi Faure, A. \& Josselin, E. 2008, \aap, 492, 25 

\noi Fields, D., Adams, N. G. \& Smith, D. 1980, \mnras, 192, 1

\noi Flower, D. R. \& Harris, G. J. 2007, \mnras, 377, 705

\noi Furlan, E. et al.~2006, \apjs, 165, 568

\noi Geers, V. C. et al.~2006, \aap, 459, 545

\noi Geers, V. C., van Dishoeck, E. F., Visser, R., Pontoppidan, K. M., 
Augereau, J.-C., Habart, E., \& Lagrange, A. M. 2007, \aap, 476, 279 

\noi Geers, V. C., van Dishoeck, E. F., Pontoppidan, K. M., Lahuis, F.,
Crapsi, A., Dullemond, C. P. \&  Blake, G. A. \aap, in press (arXiv 0812.3664)

\noi Glassgold, A. E., Feigelson, E. D. Montmerle, T. 2000, in
Protostars \& Planets IV, ed. V. Mannings, A. P. Boss, \& S. Russell
(Tucson, Univ. Arizona), p.~429

\noi Glassgold, A. E., Najita, J. \& Igea, J. 1997, \apj, 480, 344

\noi Glassgold, A. E., Najita, J. \& Igea, J. 2004, \apj, 615, 972 (GNI04)

\noi Gorti, U. \& Hollenbach, D. 2004, \apj, 613, 424

\noi Gorti, U. \& Hollenbach, D. 2008, \apj, 683, 287

\noi Goto, M. et al.~2008, \apj, in press (arXiv0811.2220) 

\noi Gougeon, S. 1998, Th\`{e}se de doctorat, Univ. Paris VII

\noi Grosso, N., Montmerle, T., Fern\'{a}ndez, Granklin, K.,  
Zapatero, O. M. R. 2007, \aap, 475, 607

\noi Habart. E. et al.~2006, \aap, 449, 106

\noi Hartmann. L., D'Alessio, P., Calvet, N. \& Muzerolle, J. 
2006 \apj, 648, 484

\noi Hollenbach, D. J. \& Salpeter, E. E. 1971, \apj 163, 155 

\noi Hornek\ae r, L. et al. 2006a, \prl, 96, 156104

\noi Hornek\ae r, L. et al. 2006b, \prl, 97, 186102

\noi Jacobs T. A., Giedt, R. R. \& Cohen, N. 1967, J. Chem. Phys. 47, 54

\noi Jewitt, D., Chizmadia, L., Grimm, R, \& Prialnik, D. 2006, in 
Protostars and Planets V, eds, B. Reipurth, D. Jewitt \& K, Keil 
(Tucson: Univ. Arizoa), p. 863

\noi Jonkheid, B., Kamp, I., Augereau, J.-C. \& van Dishoeck, E. F.
\aap, 453, 163

\noi Knez, C, et al.~2007, B.A.A.S, 39, 812

\noi Krolik, J. H. \& Kallman, T. R. 1983, \apj, 267, 610

\noi Lepp, S. \& Dalgarno, A. 1996, \aap, 306, L21

\noi Li, Z.-Y. \& Shu, F. H. 1996, \apj, 468, 261

\noi Maloney, P. R., Hollenbach, D. J. \& Tielens, A. G. G. M.
1996, \apj, 4446, 561

\noi Mandell, A., Mumma, M. J., Blake, G. A., Bonev, B. P., 
Villaneuva, G. L. \& Salyk, C. 2008, \apj, 681, L25

\noi Markwick, A. J., Ilgner, M., Millat, T. J. \& Henning, Th. 2002, \aap, 385, 632

\noi Matsuyama, I., Johnstone, D. \& Hollenbach, D. 2009, arXiv:0904.3361v1

\noi \bf \noi Meijerink, R. \&  Spaans, M. 2005, \aap, 436, 397

\noi Meijerink, R., Glassgold, A. E. \& Najita, J. R. 2008, \apj, 676, 518

\noi Meijerink, R. \&  Glassgold, A. E. 2009, in preparation

\noi Mitchell, J. B. A., Rebrion-Rowe, C., LeGarrec, J.-L., Taupier, G.
Huby. N. \& Wulff, M. 2002, \aap, 86, 743

\noi Najita, J. R., Carr, J. S., Glassgold \& Valenti, J. A.
2007, in Protostars and Planets V, eds, B. Reipurth, 
D. Jewitt \& K, Keil (Tucson: Univ. Arizoa), p. 507  

\noi Nomura H. \& Millar, T. J. 2005, \aap, 438, 923

\noi Nomura, H., Aikawa, Y., Tsujimoto, M., Nakagawa, Y. \& Millar, T. J.
2007, \apj, 661, 334

\noi Nomura, H., Aikawa, Y., Nakagawa, Y. \& Millar, T. J.2009, ApJ in press, 
(arXiv0810.4610)

\noi Perets, H. B., Biham, O., Manic\'{o}, G., Pirronello, V., Roser, J., 
Swords, S. \& Vidali, G. 2005, \apj, 627, 850

\noi Parets, H. B. et al.~2007, \apj, 661, L163

\noi Rauls, E. \& Hornek\ae r, R. L. 2008, \apj, 679, 531

\noi Salyk, C., Pontippidan, K. M., Blake, G. A., Lajuis, F., 
van Dishoeck, E. F. \& Evans, N. J. 2008, \apj, 676, L49

\noi Siebenmorgen, R. \& Kreugel, E. 2008, Heidelberg Symposium on Radiative 
Transfer, \\(www.mpia.de/RT08/par.htm)

\noi Stauber, P., Doty, S. D., van Dishoeck, E. F., \& Benz, A. O. 
2005, \aap, 440, 949

\noi Stauber, P., Jorgensen, J. J., an Dishoeck, E. F., Doty, S. D., 
and Benz, A. O. 2006, \aap, 453, 555 

\noi Tappe, A., Lada, C. J., Black, J. H. \& Muench, A. A. 2008, \apj, 680, L117

\noi Terquem, C. E. J. M. L. J. 2008, \apj, 689, 531 

\noi Turner, N. J. \& Sano, T. 2008, \apj, 679, L131

\noi van Dishoeck, E. F. \& Dalgarno, A. 1983, \jcp, 79, 873

\noi Vidali, G., Roser, J. E., Manic\'{o}, G., Pirronello, V. 2004, 
J. Geophs. Res. 109, E07S14

\noi Voit, G. M. 1992, \mnras, 258, 841

\noi Wood, K., Smith, D., Whitney, B., Stassun, K., Kenyon, S. J., 
Wolff, M. J., \& Bjorkman, K. S. 2001, \apj, 561, 299

\noi Woodall, J., Ag\'{u}ndez, M., Markwick-Kemper, A. J. \&
Millar, T. J.  2007, \aap, 466, 1197

\noi Woods, P. M. \& Willacy, K. 2007, \apj, 655, L49

\noi Woods, P. M. \& Willacy, K. 2009, \apj, in press (arXiv0812.0269)

\noi Zecho, T. et al. 2002, Chem. Phys. Letts. 2002, 366 188

\end{document}